# Toward buoyancy-driven flow at Campi Flegrei: coupled phase change and asymmetric geometry

Arturo Tozzi (corresponding author)
ASL Napoli 1 Centro, Distretto 27, Naples, Italy
Via Comunale del Principe 13/a 80145
tozziarturo@libero.it

## ABSTRACT

Bradyseism at Campi Flegrei is usually interpreted in terms of hydrothermal pressurization and magmatic degassing. Fluid flow, often treated as a passive response to pressure accumulation, is commonly modeled using simplified geometries and homogeneous permeability fields. We introduce a model in which phase transition, structural heterogeneity and geometric asymmetry jointly influence fluid flow and pressure distribution within a heterogeneous subsurface environment. We hypothesize that coupling among phase change, density gradients and flows may follow a mechanism similar to the self-propulsion observed in asymmetric floating bodies like melting ice blocks, where phase change generates buoyancy-driven currents along their inclined surfaces and net motion in the opposite direction. We simulate pressure evolution in a shallow gas-rich reservoir subject to time-dependent forcing and hydraulic relaxation, coupled to buoyancy-enhanced Darcy flow along prescribed preferential pathways. Our numerical simulations, grounded in reported deformation rates and seismicity depths at Campi Flegrei, reproduce temporal variations in uplift and the persistence of spatially localized flow. Within this framework, asymmetric geometry may promote channelized upward transport, while phase change may enhance buoyancy and contribute to pressure redistribution. Our model predicts nonlinear uplift acceleration, shallow localized seismicity and velocity scaling with pressure and buoyancy. Integration with existing multiphase models would enable the examination of how buoyancy-driven flows influence pressure evolution and deformation during volcanic unrest.

**KEYWORDS:** permeability; buoyancy; degassing; poroelasticity; convection.

## INTRODUCTION

Bradyseism at Campi Flegrei is investigated through geodetic, geochemical and geophysical observations, with a focus on hydrothermal pressurization, magmatic degassing and poroelastic deformation of the shallow crust (Acocella 2021; Picozzi and Iaccarino 2021; Di Martino et al. 2022; Astort et al., 2024; Barone et al. 2025). Ground uplift is described as the response to pressure variations within a gas-rich reservoir, often located between two- and four-kilometers depth, fed by deeper magmatic sources (Martini et al. 2008; Guidoboni and Ciuccarelli 2011; Zaccarelli and Bianco 2017; Sabbarese et al. 2020; Giacomuzzi et al. 2024; Tramelli, Tan et al. 2025). Integration of multiphase fluid flow, thermal effects and mechanical deformation provides quantitative reconstructions of recent uplift episodes and seismicity patterns. These approaches often assume isotropic and homogeneous permeability fields, treating fluid migration primarily as a diffusive or pressure-driven process (Sbrana, Marianelli, and Pasquini 2021; Convertito, and Godano 2024; Chalumeau et al. 2024; Di Toro 2025). This perspective constraints the ability to capture spatially localized deformation, transient accelerations and the organization of fluid pathways observed in seismic tomography and resistivity data. Still, phase transitions like boiling and gas exsolution are incorporated as source terms without explicitly addressing their role in structuring flow patterns. As a result, the coupling between structural asymmetry, phase change and directed fluid motion is still unresolved.

We introduce a theoretical formulation in which geometric asymmetry and phase change jointly influence fluid flow and pressure distribution. We consider a shallow reservoir subject to time-dependent input from a deeper source and to internal transformations associated with gas exsolution and boiling. Pressure evolves according to a balance between source input and hydraulic relaxation, while fluid velocity is determined by permeability, viscosity, pressure gradients and buoyancy arising from density contrasts. Coupling among phase change, density gradients and directed flow could be viewed in analogy with the melting-driven self-propulsion observed in asymmetric floating bodies, where phase change induces buoyancy-driven currents and net motion through momentum exchange with the surrounding fluid (Berhanu et al. 2026). In our setting, preferential pathways are introduced in terms of geometric features representing heterogeneous permeability and structural discontinuities. Phase change enhances buoyancy and modifies local density, amplifying upward transport along these pathways and redistributing pressure within the reservoir. We provide numerical simulations based on recent monitoring data of Campi Flegrei to explore how transient forcing, relaxation timescales and pathway geometry could influence the temporal evolution of uplift and surface deformation's localization.

METHODS

We studied the recent unrest of Campi Flegrei through a quantitative model coupling reservoir pressurization, phase change, buoyancy-enhanced flow and surface deformation within an asymmetric subsurface geometry. We used published monitoring constraints on uplift, seismicity depth, reservoir structure and degassing to parameterize a time-dependent simulation. By combining temporal evolution of pressure with spatially organized flow and a simplified mechanical response, we tested whether the interaction between phase change and imposed geometric asymmetry could reproduce both the temporal variability and spatial localization observed during recent unrest.

**Observational constraints and quantitative parameterization.** The parameter space was constrained using recent observational data (De Landro et al. 2025; Rapagnani et al. 2025). The February 2026 surveillance bulletin reported 254 earthquakes, with 192 located events concentrated within the upper 3 km and maximum depth of 4.5 km (INGV-Osservatorio Vesuviano 2026). Uplift rates were approximately $15 \pm 3$ mm month$^{-1}$ from April to early October 2025, increased to $25 \pm 3$ mm month$^{-1}$ after October 2025 and decreased to about $10 \pm 3$ mm month$^{-1}$ by early February 2026. The cumulative uplift at Rione Terra reached 162.5 cm since November 2005, with 24.5 cm since January 2025. Structural models indicate a shallow gas-rich reservoir between 2 and 4 km, a caprock between 1 and 2 km and deeper magmatic sources below 4 km. Weekly geochemical monitoring reported fumarolic temperatures near 95°C at Pisciarelli and about 173°C at Solfatara. These quantities were used to delimit physically plausible ranges for temperature, density contrast and fluid properties.

**Pressure evolution and source-term representation.** The shallow reservoir was represented through a time-dependent excess pressure $P(t)$ in MPa. Its evolution was described by the first-order differential equation

$$\frac{dP}{dt} = \frac{S(t) - P(t)}{\tau_h},$$

where $S(t)$ is an effective source term and $\tau_h$ is a hydraulic relaxation time. This formulation assumes that pressure increases due to deep input and internal phase transitions and decreases through leakage and redistribution. The source term was written as

$$S(t) = P_b + P_1 \exp\left(-\frac{t - t_0}{\tau_d}\right),$$

where $P_b$ is a background level, $P_1$ is a transient amplitude and $\tau_d$ is a decay timescale. Parameters were set to $P_b = 1.00$ MPa, $P_1 = 1.50$ MPa, $\tau_h = 4.0$ months and $\tau_d = 3.5$ months, with $t_0$ corresponding to October 2025. Numerical integration was performed using an explicit Euler scheme with time step $\Delta t = 1$ month:

$$P_{n+1} = P_n + \Delta t \frac{S_n - P_n}{\tau_h}.$$

This equation compresses multiple processes, including degassing and boiling, into a single effective forcing and does not explicitly model thermodynamic phase equilibria (Danesi et al. 2024; Hainzl, Dahm and Tramelli 2026).

**Darcy–buoyancy flow and phase-change contribution.** Fluid motion was described using a Darcy-type relation combining pressure gradient and buoyancy effects:

$$u(t) = \frac{k}{\mu}\left(\frac{P(t) \times 10^6}{L} + \Delta\rho\, g\right).$$

Here $k$ is permeability, $\mu$ is dynamic viscosity, $L$ is the vertical extent of the flow path, $\Delta\rho$ is the density contrast induced by phase change and $g$ is gravitational acceleration. The pressure term accounts for forced flow, while the buoyancy term represents the effect of vapor exsolution and heating. Values used were $k = 2 \times 10^{-14}$ m$^2$, $\mu = 1.5 \times 10^{-5}$ Pa s, $L = 3000$ m and $\Delta\rho = 250$ kg m$^{-3}$. The resulting velocity was converted from m s$^{-1}$ to m day$^{-1}$. This linear combination neglects multiphase interactions, compressibility and feedback between flow and permeability and should therefore be interpreted as a simplified representation of transport (Tan et al. 2025; De Landro et al. 2025).

**Surface deformation and calibration procedure.** Surface uplift $w(t)$ was linked to pressure through

$$\frac{dw}{dt} = aP(t) + b,$$

with $a = 0.8333$ mm month$^{-1}$ MPa$^{-1}$ and $b = 9.1667$ mm month$^{-1}$. The discrete update equation was

$$w_{n+1} = w_n + \frac{aP_n + b}{10},$$

where division by 10 converts millimeters to centimeters. The initial condition was $w(t_0) = 151.0$ cm. This relation is calibrated to reproduce observed uplift rates rather than derived from elastic theory. A secondary diagnostic for seismicity was introduced:

$$\dot{N}(t) = N_0 \exp[\beta(w(t) - w_0)],$$

with $N_0 = 18$ events month$^{-1}$ and $\beta = 0.06$ cm$^{-1}$. This variable was used to verify qualitative consistency between deformation and seismic activity (Hainzl, Dahm and Tramelli 2026; INGV-Osservatorio Vesuviano 2024).

**Geometric construction of asymmetric flow pathways.** To represent structural heterogeneity, a two-dimensional section was defined with horizontal coordinate $x \in [-3,3]$ km and depth $z \in [0,5]$ km. A curved pathway was prescribed as

$$x_c(z) = -0.6 + 0.45(5-z) - 0.06(5-z)^2,$$

with width

$$\sigma(z) = 0.28 + 0.06(5-z).$$

The velocity field was constructed as

$$V(x,z) = V_0 \exp\left[-\frac{(x - x_c(z))^2}{2\sigma(z)^2}\right]\left[0.35 + 0.65 \exp\left(-\frac{(z-2.5)^2}{3.0}\right)\right].$$

This formulation produces a localized high-velocity corridor connecting deeper and shallower regions. The geometry is imposed and represents a hypothesis about preferential pathways rather than a reconstruction derived from data (Giudicepietro et al. 2025).

**Nonlinear interaction between asymmetry and phase change.** To explore the interaction between structural asymmetry and phase change, we defined two dimensionless parameters $A$(asymmetry index) and $B$(phase-change intensity). Uplift acceleration was expressed as

$$\alpha(A,B) = C\,A^{3/2}B^2,$$

with $C = 40$ mm month$^{-2}$. This expression is not derived from physical principles but constructed to test whether nonlinear coupling produces distinct regimes. The resulting surface was evaluated over $A, B \in [0,1]$. This step provides a hypothesis-generating exploration rather than a validated prediction.

All computations were performed in Python using NumPy for numerical arrays, pandas for time handling and Matplotlib for visualization. Temporal integration covered monthly intervals from October 2025 to February 2026. Spatial fields were evaluated on regular grids. Units were maintained explicitly. The workflow consisted of parameter selection, time integration of pressure, computation of velocity and mapping of spatial distribution. Since no automated inversion, uncertainty quantification or statistical fitting was performed, our outputs should be interpreted as representative simulations illustrating possible system behavior rather than validated reconstructions.

RESULTS

Our simulated pressure increases briefly and then relaxes over the imposed hydraulic timescale. This produces an uplift rate that decreases from higher to lower values over a few months. Within the explored parameter range, the relationship between pressure and uplift rate is approximately linear, with uplift rates decreasing smoothly as pressure declines. Upward fluid velocity, computed from combined pressure gradient and buoyancy contributions, shows a monotonic dependence on pressure, confirming that both contributions act coherently under our formulation. The vertical section shows that flow is not uniformly distributed but is concentrated along a curved pathway connecting deeper and shallower regions (Figure 1). This localization persists throughout the simulated temporal evolution, indicating that spatial organization could be controlled primarily by the imposed geometry rather than by instantaneous pressure amplitude. Overall, the combined evolution of pressure decay, declining uplift rates and persistent flow localization points towards the possible coexistence of temporal variability and spatial organization within our simplified formulation.

**Interactions between geometric asymmetry and phase change.** The two-parameter exploration of geometric asymmetry and phase-change intensity provides a model-based prediction of a nonlinear response in uplift acceleration (Figure 2). For low values of either parameter, uplift acceleration is close to zero across the explored domain, indicating that neither asymmetry nor phase change alone produces significant dynamic amplification. When both parameters increase simultaneously, the system is predicted to transition to a regime characterized by rapidly increasing uplift acceleration, with values rising sharply beyond intermediate thresholds. The resulting surface exhibits a convex shape, indicating superlinear coupling between the two variables rather than additive behavior. The absence of intermediate regimes with moderate acceleration suggests that the system may organize into distinct dynamical states under our assumptions. This prediction identifies a potentially discriminable signature, i.e., a transition from gradual to accelerated deformation as both structural asymmetry and phase-change intensity increase, which can be tested against observational data.

Overall, our simulations suggest that pressure evolution, phase change and geometric asymmetry can be combined in a reduced quantitative description to assess time-dependent uplift and spatially localized flow.

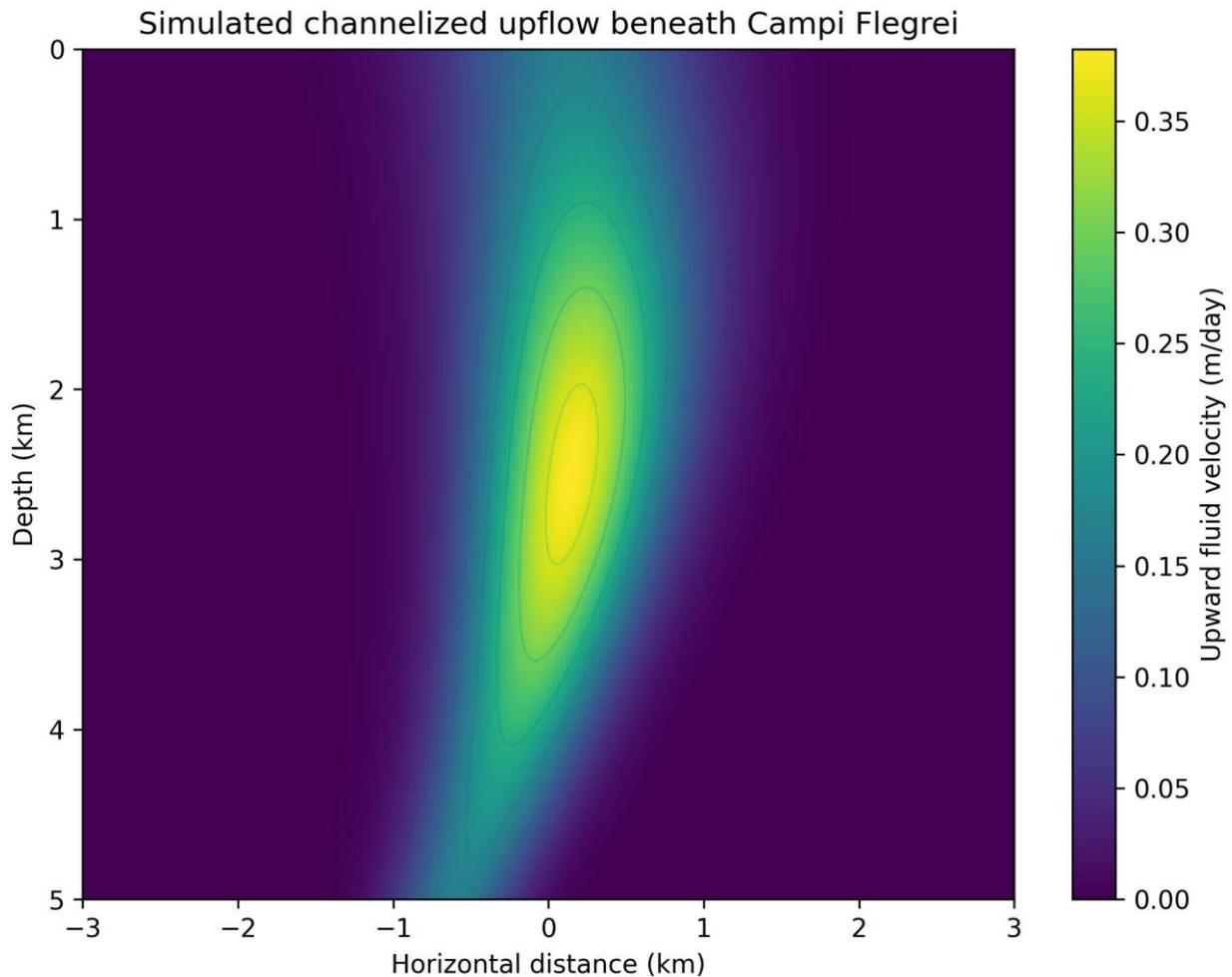

**Figure 1**. Simulated vertical section of channelized upflow beneath Campi Flegrei. The horizontal axis represents lateral distance in kilometers across the caldera, while the vertical axis indicates depth in kilometers from the surface to approximately 5 km. Color shading and superimposed contours quantify upward fluid velocity in meters per day, computed from a permeability–buoyancy formulation combining pressure gradients and density contrasts associated with phase change. The curved high-velocity corridor delineates a preferential pathway linking a deeper source region at approximately 4–5 km depth to a shallower gas-rich reservoir at approximately 2–3 km depth. The increase in velocity toward shallower depths, attributed to the combined effects of pressure-driven flow and buoyancy, is proposed as a hypothesis within our model and not directly inferred from empirical data.

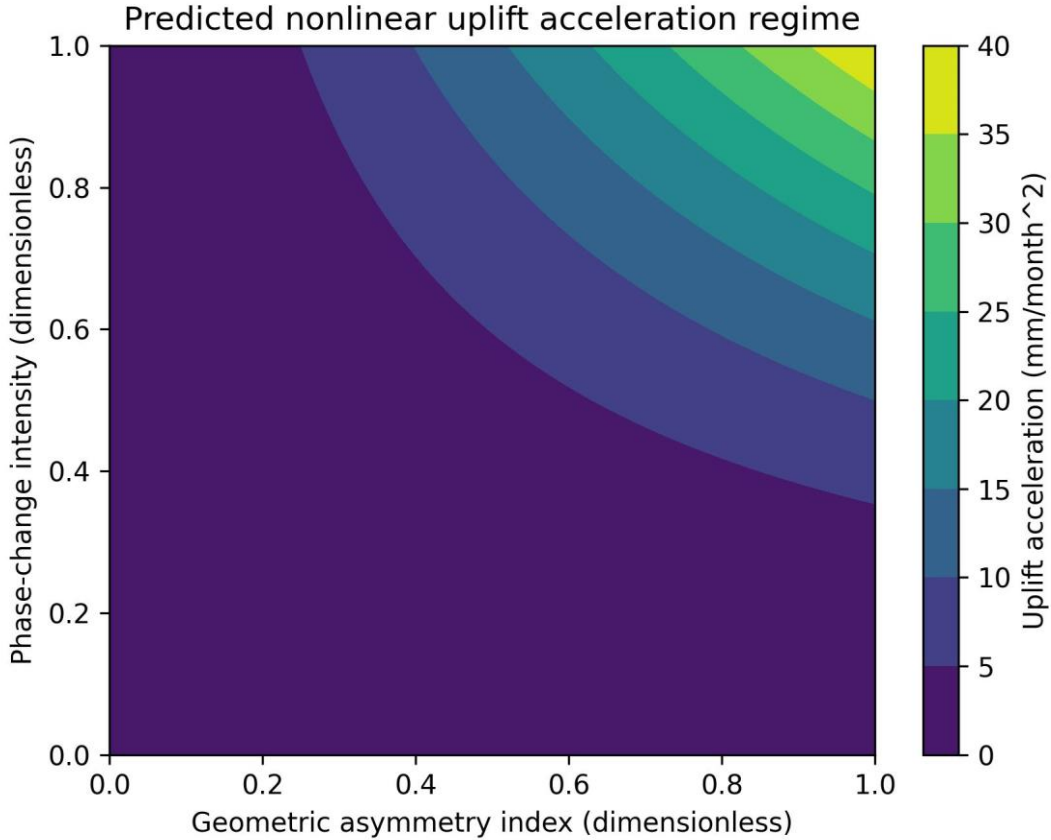

**Figure 2**. Hypothesized interaction between geometric asymmetry and phase-change intensity based on our model assumptions. Contour plot of predicted uplift acceleration (mm/month²) as a function of geometric asymmetry (dimensionless) and phase-change intensity (dimensionless), representing a model prediction. Uplift acceleration is low when either parameter is small, but increases rapidly when both exceed intermediate values. This cooperative amplification departs from additive behavior commonly assumed in hydrothermal models. The convex surface indicates a superlinear response, suggesting a transition between gradual and accelerated deformation.

CONCLUSIONS

We asked whether bradyseism at Campi Flegrei can be interpreted not only as pressure accumulation, but also as directed fluid motion arising from the interaction between phase change and geometric asymmetry. This approach is motivated by a physical observation: an asymmetric ice block melts unevenly, producing dense water that flows along its surface and generates a force that drives motion in the opposite direction. This illustrates how phase change and asymmetry can organize fluid motion. Therefore, we isolate a minimal set of variables linking flow organization to measurable quantities like uplift rate and depth distribution of activity. The coupled effect of asymmetry and phase change allows discrimination between gradual and accelerated deformation regimes within the same parameter space.
Our simulations compare time-dependent pressure evolution, buoyancy-driven flow and surface deformation within ranges constrained by monitoring data. A transient pressure increase followed by hydraulic relaxation reproduces observed uplift-rate variations, while asymmetric permeability leads to persistent localization of upward flow. In our simulations, fluid migration is organized along preferential paths connecting deeper and shallower regions rather than being uniform. The coupling between pressure gradients and density contrasts sustains coherent upward velocities even as pressure declines. In addition, the interaction between asymmetry and phase change produces nonlinear transitions in uplift dynamics. These results indicate that deformation at Campi Flegrei may reflect organized fluid transport controlled by the coupled effects of geometry and phase change

Our study has limitations. The governing equations are not derived from principles of multiphase thermo-poroelasticity. The source term $S(t)$ and the linear relation between pressure and uplift are introduced as calibrated components rather than independently validated quantities. The Darcy–buoyancy formulation assumes a linear combination of pressure gradient and density contrast, neglecting two-phase flow effects, compressibility and time-dependent permeability, all of which are relevant in the context of Campi Flegrei. The representation of geometric asymmetry relies on an imposed preferential pathway rather than one inferred from geophysical inversion. The nonlinear interaction is built through parameter choice and should be interpreted as a testable hypothesis rather than a derived law. Open questions concern how permeability evolves under repeated pressurization, how multiphase interactions modify effective density contrasts and how transient structural changes alter pathway stability.

Our model yields testable previsions. First, if phase change and geometric asymmetry jointly control deformation, then uplift acceleration should display a nonlinear dependence on indicators of fluid input and structural focusing. Specifically, the time derivative of uplift rate is expected to increase superlinearly when both gas flux and inferred pathway anisotropy rise, whereas it should remain near zero if only one factor varies.
Second, upward fluid velocity along preferential zones should scale with pressure according to a linear-plus-buoyancy relation, implying that geophysical proxies of fluid movement, such as seismic attenuation or resistivity anomalies, should increase proportionally to estimated overpressure within a defined range.
Third, the depth distribution of seismicity should remain confined within the upper three to four kilometers when flow is strongly localized, with a measurable concentration gradient aligned with the inferred pathway. These predictions can be tested through time-resolved comparisons between uplift rate, gas emissions and subsurface imaging.
Practical implications concern the integration of physically interpretable variables into monitoring strategies. The explicit computation of pathway-dependent velocity fields enables the construction of indicators sensitive to changes in subsurface transport efficiency. This may support the development of diagnostic thresholds based on combined geodetic and geochemical signals, allowing classification of unrest phases according to internal system organization rather than solely on amplitude measures. Our approach can also shape the design of monitoring networks, by identifying areas where measurements are most likely to capture early changes in subsurface dynamics. In addition, our formulation facilitates rapid updating of model states as new data become available, enabling continuous tracking of system evolution with limited computational cost.

In conclusion, we show that deformation, fluid migration and internal structure can be assessed within a unified quantitative description, albeit within a reduced representation. Our approach suggests that subsurface processes' organization, influenced by structural heterogeneity, geometric asymmetry and phase change, may jointly influence fluid migration at Campi Flegrei. This perspective provides an interpretation of temporal variability and spatial localization, while remaining compatible with existing multiphase and poroelastic descriptions.

## DECLARATIONS

**Ethics approval and consent to participate.** This research does not contain any studies with human participants or animals performed by the Author.
**Consent for publication.** The Author transfers all copyright ownership, in the event the work is published. The undersigned author warrants that the article is original, does not infringe on any copyright or other proprietary right of any third part, is not under consideration by another journal and has not been previously published.
**Availability of data and materials.** All data and materials generated or analyzed during this study are included in the manuscript. The Author had full access to all the data in the study and took responsibility for the integrity of the data and the accuracy of the data analysis.
**Disclaimer**. The views expressed are those of the author and do not necessarily reflect those of the affiliated institutions.
**Competing interests.** The Author does not have any known or potential conflict of interest including any financial, personal or other relationships with other people or organizations within three years of beginning the submitted work that could inappropriately influence or be perceived to influence their work.
**Funding.** This research did not receive any specific grant from funding agencies in the public, commercial or not-for-profit sectors.
**Acknowledgements:** none.
**Authors' contributions.** The Author performed: study concept and design, acquisition of data, analysis and interpretation of data, drafting of the manuscript, critical revision of the manuscript for important intellectual content, statistical analysis, obtained funding, administrative, technical and material support, study supervision.
**Declaration of generative AI and AI-assisted technologies in the writing process.** During the preparation of this work, the author used ChatGPT 5.3 to assist with data analysis and manuscript drafting and to improve spelling, grammar and general editing. After using this tool, the author reviewed and edited the content as needed, taking full responsibility for the content of the publication.